\documentclass[twocolumn,showpacs,preprintnumbers,amsmath,amssymb,superscriptaddress,nofootinbib]{revtex4}

\usepackage{subfigure}
\usepackage{graphicx}
\usepackage{rotating}
\usepackage{bm}

\newcommand{\rd}{\textrm{d}}
\newcommand{\ee}{\begin{equation}}
\newcommand{\eee}{\end{equation}}
\newcommand{\ea}{\begin{eqnarray}}
\newcommand{\eea}{\end{eqnarray}}

\newcommand{\msolar}{\textrm{M}_{\odot}}
\newcommand{\mbh}{\textrm{M}_{\rm H}}
\newcommand{\mbhi}{\textrm{M}_{\rm H}^{\rm ini.}}
\newcommand{\tbh}{T_{\rm H}}
\newcommand{\thawk}{T_{\rm H}}
\newcommand{\rh}{r_{\rm H}}
\newcommand{\mstar}{\textrm{M}_\star}
\newcommand{\neff}{n_{\rm eff}}
\newcommand{\nieff}{\bar n_{\rm eff}}

\newcommand{\mplanck}{\textrm{M}_{\textrm{P}}}
\newcommand{\lplanck}{l_{\textrm{P}}}

\newcommand{\gev}{\textrm{GeV}}
\newcommand{\tev}{\textrm{TeV}}

\begin{document}

\author{Michael Doran}
\affiliation{Institut f\"ur  Theoretische Physik, Philosophenweg 16, 69120 Heidelberg, Germany}
\author{J\"org J\"ackel}
\affiliation{DESY Theory Division, Notkestrasse 85, 22607 Hamburg,
Germany}

\preprint{DESY-04-251}
\preprint{HD-THEP-05-01}
\title{Testing Dark Energy and Light Particles via Black Hole Evaporation at Colliders}

\begin{abstract}
We show that collider experiments  have the potential to exclude a light
scalar field as well as generic models of modified gravity as 
dark energy candidates. 
Our mechanism uses the spectrum
radiated by black holes and can  equally well be
applied to determine the number of  light degrees of freedom.
We obtain the grey body factors for massive scalar particles and calculate the total
emissivity. While the Large Hadron Collider (LHC) may not get to the
desired accuracy, the measurement is within reach of next
generation colliders.
\end{abstract}
\maketitle

Observations indicate that our universe is in a phase of accelerated
expansion \cite{Riess:2004nr,Spergel:2003cb,Tegmark:2003ud}.  Some
mysterious dark energy seems to drive this acceleration.  Revealing
its true nature will likely entail a breakthrough in fundamental
physics.  One explanation is Einstein's cosmological constant
\cite{zeld}. It describes observations well, but is plagued by an
enormous fine tuning problem: Quantum Field Theory generically yields
a $10^{120}$ times larger value.  Scalar field dark energy cosmologies
addressing this issue \cite{Wetterich:fm,Ratra:1987rm,Caldwell:1999ew}
have been under investigation for more than a decade.  Currently,
observations only provide bounds \cite{Caldwell:2003hz,Upadhye:2004hh}
on the evolution of such an (effective
\cite{Kolda:1999wq,Doran:2002bc}) scalar field. In
addition, it may well be that the field evolution closely mimics that
of a cosmological constant in the late universe.
For years to come, astrophysical and cosmological tests may not be
able to settle the issue \cite{Kratochvil:2004gq,Upadhye:2004hh} and
perhaps may never be.  Tabletop experiments cannot be used to measure
the vacuum energy \cite{doran,Jetzer:2004vz} and hence provide no clue
to the true nature of dark energy. Likewise, direct detection of a
scalar dark energy field is next to impossible
if the interaction strength of the field is at the gravitational level
and no detectable violation of the equivalence principle is induced
\cite{Wetterich:1987fk,Sandvik:2001rv,Damour:2002nv,Parkinson:2003kf,Doran:2004ek}.
Using the cooling of an ordinary black body provides no solution
either: even though the cooling rate is proportional to the number of
degrees of freedom, scalar dark energy couples too weakly to reach
thermal equilibrium and radiate.

\begin{figure}
\includegraphics[width=0.44\textwidth]{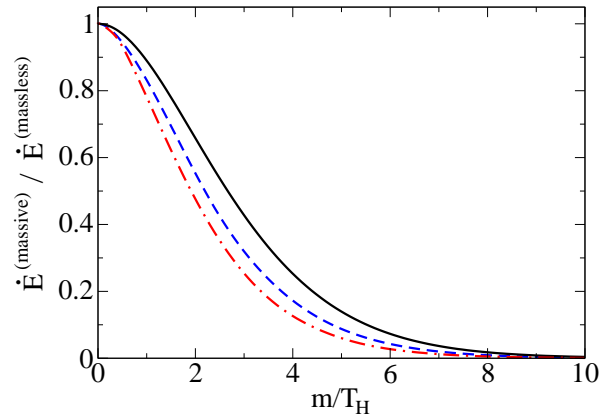}
\caption{\label{fig::emission}
Total emissivity 
of a scalar particle as a function of particle mass $m$ over
temperature $\tbh$ of a black hole with mass $\mbh \gg m$. The total emissivity
is obtained by integrating Equation \eqref{eqn::greybody} over energy $\omega$ and
using grey body factors for massive particles (see Equation \eqref{eqn::massive}).
We have normalized such that a massless scalar corresponds to unity both for 
$n=1$ dashed (blue) line and $n=7$ dashed-dotted (red) line. The normalization for $n=7$ (and
hence the un-scaled emissivity) is $\sim 136$ times larger than that for $n=1$. For comparison
we show the result for a perfect black body (solid black line).}
\end{figure}

 Here, we propose a measurement with the potential to
  exclude all scalar dark energy as well as modified gravity models.
  Our test is based on the prediction that black holes may be produced
  at colliders, provided there are extra dimensions.  By studying the
  Hawking evaporation of such black holes, it will be possible to
  count the number of light degrees of freedom -- including a scalar
  dark energy field if it exists.
Our test uses that a scalar dark energy
field has particle-like excitations with very small mass. A genuine cosmological
constant  is devoid of such excitations.
Thermodynamically  the scalar field excitation adds a
degree of freedom.  To first approximation, radiation
leaving a  black hole resembles a black body spectrum
composed of all effective degrees of freedom.
The Hawking temperature \cite{Hawking:1974sw}  corresponding
to this spectrum is given by \cite{Kanti:2004nr}
\begin{equation}
\label{eqn::temperature}
\thawk=\frac{n+1}{4\pi r_{\textrm{H}}}=\frac{n+1}{4\sqrt{\pi}}\mstar
\sqrt[n+1]{\left(\frac{\mstar}{\mbh}\right)\left(\frac{n+2}{8\Gamma\left(\frac{n+3}{2}\right)}\right)}.
\end{equation}
Here, the dimensionality of space-time is $n+4$, the Schwarzschild radius
is $\rh$, the black hole has mass $\mbh$ and the $4+n$-dimensional Planck mass $\mstar$ 
is related to the Newtonian constant by $G_{n+4} = \mstar^{-(n+2)}$. 
As the radiation originates from
gravitational interactions it is universal for all particles,
\emph{including} a scalar dark energy field (which is at least minimally coupled to
gravity).

So suppose  we knew (for details see below) the mass
of a black hole generated in a collision and also the dimensionality
of space-time. Suppose that we further knew the particle content that can
be efficiently radiated away by the black hole. From the energy deposited into
the detectors and the theoretically predicted emission of particles that
leave undetected (such as neutrinos), we can\footnote{This is only simple
in a \emph{Gedankenexperiment}. In reality the task might be a challenge to even the
finest experimental physicists.} sum up and compare
to the known mass of the black hole. A scalar dark energy field will be
excluded if there
is no energy missing.

Unfortunately, astrophysical black holes in four dimensional space-time
have temperatures
$\tbh \sim 62 \msolar / \mbh \, \textrm{nK}$, that are
too low to emit detectable radiation.
This and the (luckily) inconveniently large distance to the next black
hole make a measurement prohibitively difficult.

The way out are black holes with  temperatures
$0.1\,\textrm{eV} \lesssim \tbh \lesssim 100\,\gev$. Here, the lower limit
ensures sufficient  emission, while
the upper bound keeps the effect in an energy range where our
understanding of existing particle species is good.
If the accelerator energy is comparable to the fundamental Planck mass
$\mstar$ in higher dimensional theories
\cite{Arkani-Hamed:1998rs}, such
black holes will be produced
\cite{Argyres:1998qn,Banks:1999gd,Emparan:2000rs}.
In theories with extra dimensions, all standard model particles except for
gravitons 
are confined to a four dimensional membrane (simply called 'brane') embedded
in a higher dimensional 'bulk' space-time.
As the size $l\lesssim 1 \textrm{mm}$ of these extra dimensions may be
much larger than the typical scale $\lplanck\sim\ 1 /\mplanck \sim
10^{-32} \textrm{mm}$ of our four dimensional theory, $\mplanck\sim
\mstar^{2+n}l^n$ can substantially exceed $\mstar$. For
$n\geq 5$ higher dimensional Planck
masses as low as $\mstar\sim \tev$ are allowed by current constraints
\cite{Kanti:2004nr}.
For lower dimensions the constraints are stronger.
Roughly speaking, a black hole
forms when some energy $E \sim \mbh$ is concentrated within a radius
$\rh$.  Hence the production cross section is $\sigma\sim \pi \rh^2$,
where $\rh$ is inferred from the center of mass energy plugged into
Equation \eqref{eqn::temperature}.  Due to the smaller fundamental
Planck scale $\mstar$, higher dimensional black holes are larger (and
cooler) than their four dimensional counterparts. This
substantially increases the cross section and black holes may be
formed at reasonable rate at LHC 
\cite{Giddings:2001bu,Dimopoulos:2001hw,Harris:2004xt} or by
interactions of cosmic rays with our atmosphere
\cite{Emparan:2001kf,Feng:2001ib,Ringwald:2001vk,Anchordoqui:2001ei}

Inserting $\mstar=1\,\tev$ and $\mbh=10\,\tev$ into Equation
\eqref{eqn::temperature} we find temperature in the range
$55\,\gev-580\,\gev$ for $n=1-7$. As seen in Figure
\ref{fig::emission}, a particle can only contribute efficiently to the
evaporation for as long as its mass $m \lesssim \tbh$.  So for our
purpose, temperatures $\tbh \lesssim 580\, \gev$ are quite
\emph{agr\'eable}, in particular since future colliders may improve our
knowledge of particles up to $\sim \,\tev$. More massive black holes
are still preferable since they are cooler, emit more
particles and are less subject to quantum gravity effects.

The emission rate for one  species is
described by  \cite{Harris:2003eg}
\begin{equation}
\label{eqn::greybody}
\dot E^{(s)}(\omega)
=\sum_{j} \sigma^{(s)}_{j,n}(\omega,\rh)
\frac{\omega}{\exp(\frac{\omega}{\thawk})\pm1}\frac{\rd^{n+3}k}{(2\pi)^{n+3}}.
\end{equation}
Here $s$ and $j$ are  spin and angular momentum of the emitted
particle, $\omega = \sqrt{m^2 + k^2}$ is the
energy and $\sigma$ is the  grey body factor \cite{Kanti:2002nr,Ida:2002ez}.
In the case of
a black body, $\sigma$ is the area of the emitter. For
black holes, it is a function of the frequency of the emitted particle
which depends on the state of the black hole and in particular on the
particles mass and angular momentum. Essentially, $\sigma$ incorporates that
a particle emitted at the horizon may be reflected back into
the black hole  due to the non-trivial
interaction with the black hole. 
We have extended the calculation of \cite{Harris:2003eg} to
incorporate scalar particles of mass $m$. For these, 
the Klein-Gordon equation
in the induced black hole metric becomes
\begin{multline}\label{eqn::massive}
\frac{\rd^2  R(r)}{\rd r^2}  = -\left( \frac{2}{r} + \frac{\rd \ln[h(r)]}{\rd r} \right)  \frac{\rd R(r)}{\rd r}     \\
+ R(r) \left ( \frac{m^2}{h}  - \frac{\omega^2}{h^2}+  \frac{\lambda}{h r^2} \right ),
\end{multline}
which is to be compared to Equation (3.3) in \cite{Harris:2003eg}.
We have numerically integrated \eqref{eqn::massive} to obtain the
transmission coefficients and grey body factors 
along the lines of \cite{Harris:2003eg}.
For the massless case presented in  \cite{Harris:2003eg},
our results are in perfect agreement (i.e. we reproduced Figure 1 of
\cite{Harris:2003eg}).

Although it seems like a complication the dependence of the grey body
factors on the properties of the black hole is quite useful. In
particular, it can be used to determine the number of extra dimensions
\cite{Harris:2003eg} via the ratio of the energies emitted into
particles with different spin, i.e. scalars, gauge bosons and
fermions.

\begin{figure}[t]
\includegraphics[width=0.44\textwidth]{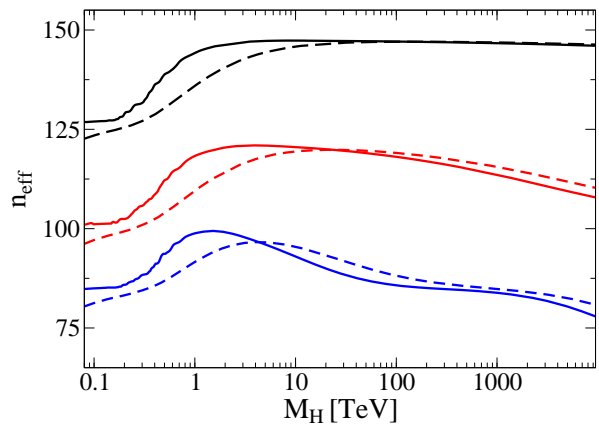}
\caption{Degrees of freedom $\neff(\mbh)$ as a function of black hole mass for 
  $n=7$ (solid upper [black] line), $n=3$ (solid middle [red] line)
  and $n=1$ (solid lower [blue] line). The dashed lines below each
  solid line are the corresponding integrated degrees of
  freedom $\nieff(\mbhi)$. Experimentally, one cannot resolve $\neff$,
  but rather measures $\nieff$.  Overall, there are more degrees of
  freedom for the $n=7$ model, because graviton polarizations
  contribute $44$ states. While the initial temperature for $n=7$ in
  the mass range depicted is sufficient to radiate all known degrees
  of freedom, the initial temperature  $\thawk^{\rm ini.} \sim 1.7\,\gev$ 
  for $n=1$ and $\mbhi = 10^4 \tev$ is too low to efficiently 
  radiate top and bottom quarks, the Higgs scalar as well as W and Z bosons.
  As the black hole evaporates, the mass drops leading to an increase in
  $\tbh$ and hence to an increase in $\neff$ until black hole masses
  $\mbh \sim 1 \tev$ are reached. At even smaller masses, $\neff$
  drops as more and more particles approach the kinematically allowed
  cut off  $\Lambda={\rm min}(\mbh[1+{m^2}/\mbh^2]/2,\mbh)$.  }
\label{fig::sm}
\end{figure}

Standard model particles and the dark energy scalar live on the brane
(of course, the dark energy scalar might also live in the bulk).  For
these, one sets $n=0$ in the integration measure of Equation
\eqref{eqn::greybody}, whereas bulk scalars and gravitons
command the full $4+n$ dimensional phase space. This does not
lead to a drastic enhancement of radiation into the bulk
\cite{Emparan:2000rs,Harris:2003eg}.  Indeed, the emitted energy per
degree of freedom for bulk fields is comparable to those on the brane.
There are, however 
$(n+3)(n+2)/2 -1$
graviton polarization states 
which for $n=7$ yields a substantial number of 
$44$ states (see also Figure \ref{fig::sm}).

The higher dimensional Planck mass can be determined from the
production cross section of gravitons in collisions where no black
hole is formed \cite{Giudice:1998ck}.  As the grey
body factors depend on the number of extra dimensions, we can furthermore
infer $n$ from the relative abundances \cite{Harris:2003eg} of
particles with different spin.  Measuring the spectrum of 
particles emitted and using Equations \eqref{eqn::temperature} and
\eqref{eqn::greybody} one can infer the radius, temperature and mass 
of the black hole.

We define the effectiveness $n^{(x)}$ of some degree of freedom $x$ 
by comparing the emission rate into channel $x$ to the
emission into one massless scalar 
\newcommand{\coff}{\Lambda}
\begin{equation}\label{eqn::defnx}
 n^{(x)}(\mbh) \equiv \int_{m}^{\coff}\rd \omega\,\dot E^{(x)}(\omega)  \Bigg /   \int_0^{\mbh/2}\!\!\!\!\rd \omega\, \dot E^{(m.s.)}(\omega) .
\end{equation}
Here, the cut-off $\Lambda={\rm min}(\mbh[1+{m^2}/\mbh^2]/2,\mbh)$ limits
the energy of emitted particles and is due to 
energy-momentum conservation and finite black hole mass
and $\Lambda \geq m$ is understood.
 Please note that
$\mbh$ decreases steadily during evaporation.  
The number of effective degrees of freedom is then given by $\neff(\mbh) = \sum_{x}
n^{(x)}(\mbh)$. It is not
directly observable, as experiments lack  resolution to connect
particles to their corresponding emission times. 
What we \emph{can}  observe is the integral over the evaporation process,
where the energy deposited into one massless scalar is 
\begin{equation} \label{eqn::emassless}
E^{(\rm m.s.)}(\mbhi) = \int^{\mbhi}_{0} \frac{\rd M}{ \neff(M)}.
\end{equation}
Inverting this relation \eqref{eqn::emassless} and normalizing to
$\mbhi$, the  integrated number of degrees of freedom $\nieff(\mbhi) \equiv \mbhi /
E^{(\rm m.s.)}(\mbhi)$ follows. In contrast to $\neff$, we can
measure $\nieff$ from the total energy $E^{(x)}$ deposited into a known species $x$ 
using  $E^{(m.s.)}=E^{(x)}/\int^{\mbhi}_{0} n^{(x)}(M)\,\rd M$ from Equation \eqref{eqn::defnx}.

If standard model particles and gravitational polarization states can\footnote{There is one subtlety concerning the still
  unknown nature of neutrinos. Dirac neutrinos will effectively
  contribute twice as much degrees of freedom as Majorana neutrinos.
  Turned around this might also give us a hint about the true nature
  of neutrinos. Nevertheless, it is quite likely that the nature of
  neutrinos can be inferred from other experiments like, e.g., ones  to
  detect the neutrinoless double $\beta$ decay.}
account for $\nieff$,  a scalar dark energy field \emph{will be ruled
  out}. The same is true for bulk scalars and weakly interacting brane
particles with masses $\lesssim \thawk^{\rm ini.}$.  This would leave
us with a cosmological constant as the only explanation for the
acceleration of our Universe.    Please note
that long distance modifications such as \cite{Carroll:2003wy} would
also be ruled out as they are equivalent to scalar field models with
light scalars \cite{teytou,Wands:1993uu,Chiba:2003ir}. 

If, on the other hand we find missing energy which cannot be accounted
for then possible candidates must have $m \lesssim \tbh^{\rm ini.}$.
Distinguishing between bulk and brane fields would then require a high
precision measurement making use of the slightly different emission 
rates.

Standard model particles contribute roughly one hundred degrees of
freedom.  In addition, we have $(n+3)(n+2)/2-1$ gravitational modes.
Assuming that the latter 
radiate approximately like 
a scalar field,
 we see that $\nieff$ needs to be determined to better than $0.5\%$ (see also Figure \ref{fig::sm}). 
 A recent study of possible black hole decays at LHC
\cite{Harris:2004xt} predicts an accuracy for the measurement of the
total energy emitted into known particles of $\sim30\%$ for a
$5\textrm{TeV}$ and $\sim15\%$ for a $8\textrm{TeV}$ black hole
($\mstar=\textrm{1TeV}$).

This is not yet sufficient for our  measurement -- but only by
two orders of magnitude. 
Future colliders will probe higher and
higher energies and produce black holes with ever increasing mass. 
As more massive black holes are cooler, they emit a smaller
variety of particles with considerably better statistics.  Hence, the
measurement proposed is within reach of next generation colliders.

Beside experimental challenges there remain theoretical problems to be
solved. First, the emission of gravity modes into the
bulk must be better understood.  Moreover, the data for Figures
\ref{fig::emission} and \ref{fig::sm} was inferred for a spherically
symmetric black hole in semiclassical approximation. Yet, a typical
black hole produced in high energy collisions goes through several
phases \cite{Giddings:2001bu}.  Initially, such a black holes is
asymmetric and has non-vanishing angular momentum as well as charges
originating from the producing particles. In the ``balding'' phase it
loses quantum numbers and asymmetry inherited from the original
collision. During the ''spin down'' phase it radiates away angular
momentum.  Then it enters the ''Schwarzschild'' phase where the black
hole is spherical and our semiclassical considerations are valid.
Finally, it enters the ''Planck'' phase where its mass is $\sim
\mstar$ and quantum gravity effects become important.  So far, only
the Schwarzschild phase in which it roughly deposits $\sim 60\%$ of
its energy is well understood. Early attempts to calculate the
''spin-down'' are underway \cite{Ida:2002ez}.

{\it Conclusions:} We have shown how missing energy in the decay of
higher dimensional black holes produced at colliders may be used to
discern the number of light particles/fields.   In particular, a
scalar dark energy field can be excluded provided all energy radiated
away from the black hole is accounted for by known particles and
graviton polarization states. Counting light degrees
of freedom could answer additional questions. It might, for example,
reveal the Majorana/Dirac nature of neutrinos.
The proposed measurement is challenging
for experimentalists and necessitates a better understanding of black
holes produced at colliders.  Yet, it may be the one and only way to
rule out a light scalar field or modified gravity as dark energy
candidates.

{\bf Acknowledgments} We would like to thank R. R. Caldwell, A. Ringwald 
and C. Wetterich  for discussions.


\end{document}